\documentstyle[editedvolume]{jd}

\newcommand{\etal}{\textit{et~al.}}
\begin{opening}
\title{BINARIES IN GLOBULAR CLUSTERS}

\author{S. L. W. McMillan}
\institute{Drexel University\\
	   Department of Physics, Philadelphia, PA 19104, U.S.A.}

\author{C. Pryor}
\institute{Rutgers, the State University of New Jersey\\
	   Dept.\ of Physics and Astronomy, 136 Frelinghuysen Rd.,
	   Piscataway, NJ 08854, U.S.A.}

\author{E. S. Phinney}
\institute{California Institute of Technology\\
           Theoretical Astrophysics, 130-33 Caltech,
	   Pasadena, CA 91125, U.S.A.}

\end{opening}

\begin{document}


%
%
\newcommand
	\simlt{\hbox{\rlap{\raise0.425ex\hbox{$<$}}\lower0.65ex\hbox{$\sim$}}}
\newcommand
	\simgt{\hbox{\rlap{\raise0.425ex\hbox{$>$}}\lower0.65ex\hbox{$\sim$}}}

\section{Dynamics of Binary Stars \textrm{(SLWM)}}

Binary stars in a globular cluster (hereafter, GC) may be primordial
(i.e.~formed along with the cluster), or the result of cluster
dynamics.  ``Dynamical'' binaries can result from conservative
three-body encounters (e.g.~Spitzer, 1987) if a third star can carry
away enough kinetic energy to leave two others bound, or from
dissipative two-body encounters, if two stars happen to pass within a
few stellar radii of one other (Fabian, Pringle, \& Rees, 1975).  Such
non-primordial systems are likely to be found primarily in evolved GC
cores, both because conditions are more favorable for making them
there, and because of mass segregation.  Knowledge of the formation
process allows reasonable estimates to be made of their mass and
energy distributions.  The initial spatial, mass, and energy
distributions of primordial binaries, on the other hand, are largely
unknown.

Until the late 1980s, the conventional wisdom was that GCs were born
with few, if any, binary systems, so GC binaries had to form
dynamically and consequently were rare.  However, there is now strong
evidence that many GCs contain binary fractions in the $\sim3$--30\%
range (see \S2), and it seems reasonable to suppose that most, and
quite possibly all, GCs began their lives containing significant
numbers ($\simgt\,10$\%, say) of primordial binaries.

\paragraph{BINARIES and GLOBULAR CLUSTER EVOLUTION.}
The role of binaries as energy sources driving GC evolution is well
known (see, e.g.~Heggie, 1975).  Briefly, hard binaries (i.e., systems
having binding energies exceeding the mean kinetic energy ${3 \over 2}
kT$ of cluster stars) tend to become more tightly bound following
interactions with other cluster members.  For equal-mass systems, the
median fractional energy increase is $\delta E/E\sim20$\%; the mean
time between interactions scales inversely with binary energy $E$, so
the time-averaged heating rate stays constant, at about $1kT$ per
local relaxation time.  The details depend somewhat on the masses of
the stars involved---in particular, the energy release shows
systematic trends with both the binary mass ratio and the mass of the
incoming star---but the overall energetics are clear.

Numerical simulations have amply demonstrated how primordial binaries,
if present at more than the $\sim5$\% level, rapidly segregate to the
cluster core and dominate both the core mass and the cluster evolution
(McMillan, Hut, \& Makino, 1990, 1991; Gao \etal, 1991; Heggie \&
Aarseth, 1992).  Even a small (1--2\%) initial population of hard
binaries can give rise to a core binary fraction in the 10--20\% range
within a half-mass relaxation time (a few billion years in typical
cases).  Subsequently, binaries control the cluster dynamics until
they are all destroyed by interactions with other binaries, or recoil
out of the cluster after a triple or four-body encounter.

The rate at which binaries interact and are depleted in the core is
largely independent of the rate at which Galactic tides strip mass
 from the cluster's outer regions.  As a result, there is a
``watershed'' initial binary fraction (of around 10--15\% for the
simple models considered so far), below which the cluster binary
fraction reaches zero before the cluster dissolves, and above which
the binary fraction first falls, then rises prior to eventual
dissolution (McMillan \& Hut, 1994).  Even if their initial binary
fraction is below this watershed value, however, the binary depletion
time for many clusters may exceed the age of the universe.

Primordial binaries stabilize the core against collapse at a radius of
a few percent of the half-mass radius.  As a result, it seems quite
unlikely that a GC will ever reach densities high enough for 3-body
dynamical binaries to form, although 2-body tidal binary formation
remains a possibility.  Monte-Carlo simulations by Hut, McMillan \&
Romani (1992) and Sigurdsson \& Phinney (1995) indicate that, at any
instant, some non-negligible fraction of cluster binaries will be
found well beyond the half-mass radius, drifting back toward the core
after an interaction whose recoil did not quite eject them from the
cluster's potential well.

It is well established that mass and angular momentum transfer in
isolated binaries can lead to evolutionary pathways not accessible to
single stars.  When interactions with other stars enter the mix, still
more possibilities arise.  ``Resonant'' encounters with other stars or
binaries almost invariably lead to very close approaches between some
pair of stars, so the presence of binaries in a cluster can greatly
increase the probability of stellar collisions and close encounters,
particularly in low-density systems (Verbunt \& Hut, 1987).  Even
without physical collisions, binary interactions can profoundly affect
binary evolution.  ``Flyby'' encounters may radically change a
binary's orbital parameters, while ``exchange'' interactions, by
preferentially ejecting the lightest of the stars involved, can lead
to systematic and largely irreversible changes in overall binary
composition.

\paragraph{SIMULATIONS of BINARY-RICH STAR CLUSTERS.}
The presence of a substantial primordial binary population leads to an
intimate coupling of cluster dynamics and stellar evolution.  Binaries
control the dynamics, but the cluster environment strongly influences
each binary's evolution and survival probability.  Over time, the most
massive objects in a cluster tend to find their way into binaries in
or near the core; the binaries then mediate interactions between these
objects, possibly producing many of the exotica observed in GCs today
(see \S3).  Perhaps surprisingly, given the important connection
between observations and our understanding of cluster dynamics,
theoretical simulations of cluster dynamics have tended to stop short
of including stellar-evolutionary effects.  To date, {\it no} fully
(or even nearly) self-consistent calculation including both dynamics
and evolution has ever been carried out.

One reason for this is the ``kitchen sink'' effect that follows when
one contemplates moving beyond the simplest simulations.  For the most
part, model systems incorporating binaries have consisted of
identical, non-evolving point masses---a good starting point, but
hardly an accurate model of any real star cluster.  Such a simple
system has only one relevant parameter---the total number of stars, N.
However, when we add a spectrum of stellar masses, we necessarily
introduce real stellar physics into the calculation---a stellar mass
function must be chosen, and the spatial distribution of each stellar
species defined.  Once this is done, it immediately becomes necessary
also to include the effects of stellar evolution---stellar
evolutionary time scales are often comparable to the relaxation/core
collapse time of the cluster, so the dynamical state of the cluster is
necessarily coupled to the state of the component stars.  Stellar
evolution drives binary evolution, so this too must be incorporated if
the simulation is to remain self-consistent, and the seemingly
innocuous act of including stellar masses actually leads inevitably to
a complex mix of physical processes and a vastly more difficult
numerical problem.

Over the past few years, a number of groups have been gearing up to
meet this challenge.  It seems clear that ``pure'' large-N techniques,
such as Fokker-Planck and gas-sphere codes, are structurally incapable
of following systems containing large numbers of evolving binaries and
evolution products.  However, when coupled with Monte-Carlo treatments
of binaries (see Spurzem, these proceedings), these methods may regain
applicability to binary-rich systems.  Giersz (these proceedings) has
recently reported significant progress in full Monte-Carlo cluster
simulations, with and without binary populations.  On the N-body side,
Aarseth's venerable NBODY4 (Aarseth, 1996) has recently been joined by
the ``kira'' integrator developed by McMillan, Hut, Makino, and
Portegies Zwart as part of the Starlab software environment (see
McMillan \& Hut, 1996).  Both NBODY4 and kira include sophisticated
treatments of stellar and binary evolution as integral parts of the
programs (see Tout and Portegies Zwart, these proceedings; also
Portegies Zwart et al, 1998).

%
%
\section{Observations of Binary Stars \textrm{(CP)}}
This section briefly discusses some recent results constraining the
number and properties of binary stars in GCs.  More complete reviews of
this field are Hut \etal\ (1992; H92 hereafter), Pryor \etal\ (1996),
and Meylan \& Heggie (1997).  Most of the information on the GC binary
population comes from systematic surveys for eclipsing binaries, for
radial velocity variables, and for stars above the main sequence in the
color-magnitude diagram (CMD), also called ``binary sequences.''  See
H92 for an extensive discussion of the techniques used in these
surveys.  Exotic stellar objects in GCs, which are discussed in \S 3,
are powerful probes of the interplay between binaries and GC dynamical
evolution, but are less useful as guides to the total binary
population.

Eclipsing binaries are rare in GCs, but present (see Mateo, 1995 and
\S 9.6 of Meylan \& Heggie, 1997 for reviews).  The frequency of these
close binaries is probably an interesting probe of GC dynamics, but a
poorer guide to the total binary population.  Contact (W~UMa) systems
are the most easily detected and primordial binaries with periods
shorter than perhaps five days are expected to have been brought into
contact over the age of the GCs by the angular momentum loss associated
with magnetized winds (Eggen \& Iben, 1989).  Unfortunately, the
frequency of W UMa systems is difficult to compare to that of the other
binary populations because W~UMa lifetimes are uncertain by about a
factor of 100 (Eggen \& Iben, 1989).  Indeed, it now probably more
fruitful to use GC binary frequencies to constrain these lifetimes
rather than {\it vice versa}.

Most surveys for radial velocity variables in GCs have studied luminous
giants and only been sensitive to orbital periods ($P$s) in the two-decade
range between 0.2 and 20~years.  The lower limit comes from
the smallest orbit in which the giants can fit without Roche-lobe
overflow and the upper limit from the time span of the observations and
their accuracy.  These studies find that an average of 0.07
of the stars in GCs are the primary of a binary with a period
in an arbitrary decade within the above range and a mass ratio larger
than 0.22 (see Pryor \etal, 1996 and H92 for the details).  Statistical
and systematic uncertainties allow this fraction to be 0.03--0.15.

This GC binary frequency can be compared to the value of 0.06 per
decade of period for $P \approx 2$~yr found for local solar-type stars
(Duquennoy \& Mayor, 1991).  It is surprising that GCs are not obviously
deficient in binaries compared to the Population~I field, given that
binaries are expected to be destroyed by dynamical processes in GCs
(Heggie, 1975, H92, \S 1).  However, the period ranges that are most
vulnerable to elimination are those with $P\, \simlt\, 5$~days, as
noted above, and those with
\begin{equation}
P\, \simgt\, (65\ {\rm yr})\left(5\ {\rm km~s}^{-1}/\sigma\right)^3,
\end{equation}
where $\sigma$ is the cluster velocity dispersion (e.g. Pryor
\etal, 1996).  These long-period binaries have binding energies smaller
than the typical kinetic energy of stars in the cluster and are
expected to be quickly disrupted (Heggie, 1975).  The cluster must also
be sufficiently dense for all of these ``soft'' binaries to have
suffered an encounter; see, for example, Pryor \etal\ (1996).

Recent radial-velocity surveys have increased the range of orbital
periods that can be detected either by increasing the time baseline or
precision of the measurements (Mayor \etal, 1996 -- $\omega$~Cen;
C\^{o}t\'{e} \etal, 1996 -- M22) or by observing less luminous
giants (Barden \etal, 1995 -- M71; Yan \& Cohen, 1996 -- NGC~5053)
or actual main-sequence stars (C\^{o}t\'{e} \& Fischer, 1996 -- M4).
The searches for short-period binaries found binary frequencies
comparable to or larger than the previous average value.  The searches for
long-period binaries in $\omega$~Cen and M22 found smaller values.
This led C\^{o}t\'{e} \etal\ (1996) to suggest that the expected
destruction of long-period binaries had been detected.  This seems
likely, but larger surveys in individual GCs and a better knowledge of
the orbital ellipticity distribution (an important systematic
uncertainty, see H92) are needed to be certain.

A possible reason why the present-day frequency of hard binaries in GCs
is not clearly below that of the Pop~I field is provided by the
increasing evidence for enhanced binary frequencies in GCs which have
lost a lot of mass to the galactic tidal field.  As discussed in \S 1,
mass segregation and tidal mass loss can increase the binary frequency
(McMillan \& Hut, 1994).  The most spectacular confirmation of this
prediction is the \textit{open} cluster NGC~3680, for which a careful
radial velocity and proper motion survey found no main-sequence members
fainter than 1.5~mag below the turnoff (Nordstr\"{o}m \& Andersen
1997).  This suggests that 90\% or more of the initial cluster mass has
been lost for reasonable initial mass functions.  Over 50\% of cluster
members are found to be binaries, a much larger fraction than has been
found by comparable surveys of Population~I field stars.

Among the GCs, M71 has a low total mass and an orbit confined to the
galactic disk, suggesting that it has suffered extensive tidal mass
loss, and has a binary fraction at the upper end of the observed range
(see Pryor \etal, 1989 and C\^{o}t\'{e} \etal, 1996).  Veronesi
\etal\ (1996) estimate that the very low luminosity cluster E3 has
a binary frequency (including all orbital periods) of about 30\% based
on a binary sequence in the CMD.  Clearly, it will be interesting to
look for a correlation between binary frequency and indicators of the
amount of tidal mass loss.

Searches for binary sequences benefit from the high angular resolution
of the Hubble Space Telescope (HST), which reduces the number of
coalesced stellar images, caused by crowding, that mimic binaries (see
Romani \& Weinberg, 1991, H92).  Rubenstein \& Bailyn (1997) studied
the post-core-collapse (PCC) GC NGC~6752 and found a binary frequency
of 15--38\% (99.7\% confidence interval) within 11$^{\prime\prime}$ of
the center and 0--16\% beyond.  These frequencies include binaries
with all periods that are present.  These numbers and the CMDs
themselves suggest that the binaries are strongly concentrated towards
the cluster center, as would be expected from mass segregation (see \S
1).  What is surprising is that there is such a large population of
binaries near the cluster center at all.  The projected radius
containing half of the cluster light is about 140$^{\prime\prime}$
(Harris, 1996), so, as discussed in \S 1, the binary population
probably should have maintained a resolvable core.  Detailed dynamical
modeling of NGC~6752 would be interesting.  Searches for binary
sequences also need to carried out in the centers of other PCC GCs.

Fahlman \etal\ (1997) have estimated the binary frequency in the nearby
non-PCC GC M4 using the CMD derived from HST data extending from the
center out to about 6 core radii.  The binary frequency is 0.05 with a
statistical uncertainty of $\pm$30\% (2$\sigma$) and there is some
evidence that the frequency increases inward.  This value is no higher
and probably lower than the average GC binary frequency from radial
velocity surveys, despite being sensitive to a much wider range of
periods.  This perhaps suggests that a narrower range of binary periods
has survived than predicted by theory.  An extensive radial velocity
survey is underway in M4.  A prediction of the properties of the binary
population expected to have survived in the cluster would also be
useful.

%
%
\section{Exotic Stars and Binaries in Globular Clusters
\textrm{(ESP)}}

The mass function of non-degenerate stars in GCs now extends only to
$\sim 0.8M_\odot$.  However the dynamical and chemical evolution of
clusters depends crucially on much more massive stars now present only
as cold relics: black holes, neutron stars and white dwarfs.  These
are detectable only through their gravitational effects unless lit up
by accretion from a binary companion.  In a bright cluster of $L_V\sim
10^6L_{V\odot}$ (e.g.~M15, 47~Tuc), extending a Salpeter initial mass
function (IMF) gives 5000 neutron star progenitors ($12-25M_\odot$),
2000 black hole progenitors ($25-60M_\odot$), and $\sim 3\times 10^5$
white dwarfs.  The whole GC system has about 20 times this luminosity,
about $10^{-3}$ of the luminosity of the Milky Way.  Yet the clusters
contain $10^{-1}$ of the low-mass X-ray binaries in the Galaxy (the
same ratio applies to the X-ray sources in M31 and its GCs,
\cite{supper97}), and about half of the known recycled pulsars.  In
contrast, there are no black hole candidates in clusters, and
cataclysmic variables have been hard to find, and are clearly not so
overabundant as X-ray binaries and recycled pulsars.

These systematics, and the properties of the binaries (particularly
the pulsars) are determined by binary interactions and their
back reaction on cluster evolution.  These exotic binaries still
provide our main observational evidence for binary interactions and
their importance in cluster evolution.

\paragraph{BLACK HOLES.}
There is no evidence for black holes in GCs.
The radial distribution of pulsars in the center of M15 limits
the number of $\sim 10M_\odot$ black holes in that cluster to 
$<100$ \cite{phinney93}. Thermonuclear bursts
 from all ten of the well-studied X-ray sources in GCs
rule out the possibility that they are accreting black holes.
This is at first quite surprising.
If stars more massive than $25M_\odot$ create black holes, 
then for every neutron star, 0.4 black hole should be created
in an $x=1.35$ Salpeter IMF (0.1 would be created for $x=3$).
The retention fraction, dynamical friction and exchange cross-section
should all be larger than for neutron stars. 
So why have no black hole X-ray binaries been seen?  
A likely explanation is that the clusters were born with
many black holes, but that these rapidly segregated to the cluster cores,
formed binaries, and ejected themselves by binary 
interactions \cite{sigurdsson93,kulkarni93}.
A few could be left, particularly in lower density clusters, and
might appear as X-ray transients.

\paragraph{NEUTRON STARS and WHITE DWARFS.}
The properties of cluster X-ray binaries have been reviewed by Verbunt
\shortcite{verbunt96} and Bailyn \shortcite{bailyn96} and those of
cluster radio pulsars by Phinney \shortcite{phinney96}.  At birth,
neutron stars are kicked to a velocity of $\sim 300\,\mbox{km s}^{-1}$
\cite{hansen97}, much higher than the escape velocity from GCs.  How
were cluster neutron stars retained for recycling?  Perhaps from a
low-velocity tail of the kick distribution, but binaries are probably
the main source.  If the more massive star in a close circular binary
goes supernova (the progenitor of a low-mass X-ray binary [LMXB]), the
binary will remain bound only if the remnant neutron star is kicked
with almost the same speed and direction as the low-mass companion.
Thus the center of mass velocity will be about equal to the (high)
pre-supernova relative orbital velocity, explaining why Galactic LMXBs
and low-mass binary pulsars are rare and fast-moving. But if the less
massive star goes supernova (typical of high-mass X-ray binaries,
where the initially more massive star undergoes unstable mass
transfer, becomes the less massive one, and explodes as a helium
star), the binary will remain bound mainly if the kick is about equal
and opposite to the exploding star's orbital velocity (double and
opposite flips the orbit over).  The final center of mass velocity
will be approximately the (low) pre-supernova orbital velocity of the
massive companion.  The subset of such systems in which there is no
second spiral-in and supernova are the ones most likely to dominate
neutron star retention in clusters.  Specific models have been
computed \cite{brandt95,kalogera96,drukier96}.

Today, the retained neutron stars are the most massive stars in GCs,
so they tend to sink to the dense cores and to be preferentially left
in binaries in 3 and 4 body exchanges \cite{sigurdsson95,heggie96}.
This is why X-ray binaries and recycled pulsars are so overabundant in
clusters relative to the Galactic disk.  Though white dwarfs are much
more numerous, they are less massive, so less centrally concentrated,
and less likely to be retained in binary exchanges.  This is probably
why cataclysmic variables (CVs) in clusters are not nearly so
overabundant in clusters, though their numbers and formation
mechanisms \cite{davies97} are still poorly understood.  Massive white
dwarfs, which in the Galaxy seem to be preferentially magnetic
\cite{sion88}, ought to be overrepresented.

The next few years should
show great growth in the numbers of exotic objects available for
study.  Since 1994 the number of CV candidates in GCs has risen
 from 2 to 13:
NGC6397 [3] \cite{cool95,grindlay95}, 47~Tuc [4] 
\cite{paresce94,edmonds96,minniti97,shara97}, M4 [1] \cite{kaluzny97},
M5 [1] \cite{hakala97}, NGC6624 [1] \cite{shara96}, NGC 6752 [2] 
\cite{bailynea96}, M80 [1] \cite{shara95}.
With the refurbished Arecibo, GBT and GMRT coming on line, the number
of radio pulsars should grow dramatically.  The increased angular
resolution of AXAF should help resolve questions about the nature of
clusters' low luminosity X-ray sources.

\paragraph{STARS.}
Nondegenerate stars are not immune from exchanges and two and three-body tidal
encounters.  Stars that have been proposed as candidate victims include
high velocity stars in 47~Tuc \cite{meylan91} [exchange],
bright blue stragglers \cite{bailyn95,dantona95} [collisional mixing],
rapidly rotating
horizontal branch stars \cite{peterson95,cohen97} [mergers], extended blue
horizontal branch stars \cite{sosin97,rich97} [envelope heating
or stripping], and the missing
red giants in the cores of core-collapse clusters 
\cite{djorgovski91,burgarella96} [envelope stripping], though the numbers
and radial distributions of objects cause difficulties for
many of these explanations.

%
%
\bigskip\noindent\textbf{Acknowledgments:}~~SLWM's research has been
supported by NASA grant NAGW--2559 and NSF grant AST 93--08005.  CP's
research on GCs is supported by NSF grant AST 96--19510. ESP's research
has been supported by NASA grant NAG 5--2756 and NSF grant AST 96--18537.

{}

\end{document}